\documentclass[letter]{aa}  

\usepackage{graphicx}
\usepackage{txfonts}
\usepackage{color}
\definecolor{jf}{rgb}{1.0,0.6,0.}

\definecolor{mhi}{rgb}{0.6,0,0.6}

\definecolor{aza}{rgb}{0.0,0.5,0.0}

\newcommand{\hi } {{\rm H}\,{\small\rm I}}

\begin{document} 
   \title{A tidal tale: detection of multiple stellar streams in the environment of NGC\,1052
   }
 \titlerunning{Deep HERON imaging of the NGC1052 field}
   \author{Oliver M\"uller\inst{1}
           \and
          R. Michael Rich\inst{2}
\and
Javier Rom\'an\inst{3}\inst{4}
          \and
          Mustafa K. Y{\i}ld{\i}z\inst{1}\inst{5}\inst{6}
          \and
          Michal B\'ilek\inst{1}
        \and
          Pierre-Alain Duc\inst{1}
                    \and
          J\'er\'emy Fensch\inst{7}
          \and 
          Ignacio Trujillo\inst{3}\inst{4}
          \and
                    Andreas Koch\inst{8}
          %
          %
          %
}
   \institute{Universit\'e de Strasbourg, Observatoire Astronomique de Strasbourg  (ObAS), CNRS UMR 7550 Strasbourg, France \\\email{oliver.muller@astro.unistra.fr}
   \and  
   Dept. of Physics and Astronomy, UCLA, Los Angeles, CA 90095-1547
   \and 
Instituto de Astrof\'isica de Canarias (IAC), La Laguna, 38205, Spain
\and
Departamento de Astrof\'isica, Universidad de La Laguna (ULL), E-38200, La Laguna, Spain
   \and
   Astronomy and Space Sciences Department, Science Faculty, Erciyes University, Kayseri, 38039 Turkey
          \and
         Erciyes University, Astronomy and Space Sciences Observatory Applied and Research Center (UZAYB\.{I}MER), 38039, Kayseri, Turkey
   \and
   European Southern Observatory, Karl-Schwarzschild Strasse 2, 85748, Garching, Germany
   \and
    Astronomisches Rechen-Institut, Zentrum f\"ur Astronomie der Universit\"at Heidelberg, M\"onchhofstr.\ 12--14, 69120 Heidelberg, Germany
   }
   \date{Received tba; accepted tba}

 
  \abstract
 {The possible existence of two dark matter free galaxies (NGC1052-DF2 and NGC1052-DF4) in the field of the early-type galaxy NGC\,1052 {presents} a challenge to theories of dwarf galaxy formation according to the current cosmological paradigm.
 We carried out a search for signatures of past interactions connected to the putative hosts of NGC\,1052-DF2 and  NGC\,1052-DF4 using a very deep {$L$-band} image obtained with the 0.7\,m Jeanne Rich telescope that reach a surface brightness limit of 28.5\,mag arcsec$^{-2}$ in the $r$ band. 
 We found several low-surface brightness features, possibly consistent with an ongoing merger history in this group. We find 
 {a tidal interaction between}
 NGC\,1052 and NGC\,1047, 
 confirming a physical association.  Furthermore we find a stellar loop around NGC\,1052 in the direction of NGC\,1042 and a stellar stream pointing in the direction of NGC\,1052-DF2 -- but both are not directly connected.
 We find no evidence for a recent tidal interaction for NGC\,1052-DF2 and NGC\,1052-DF4.
 No low surface brightness features have been uncovered around the   spiral galaxy NGC\,1042,  
 { which leaves the association (physical or projected) between NGC\,1052 and NGC\,1042 ambiguous, though they have similar radial velocities. Their association  will only be established once  accurate distances to both objects have been measured.
 }
 }

   \maketitle
%

\section{Introduction}
The discovery of 
two apparently dark matter free dwarf galaxies in the field of the early-type galaxy NGC\,1052 has {sparked strong interest with numerous follow-up studies} \citep[e.g. ][]{2018Natur.555..629V,2018ApJ...864L..18V,2019MNRAS.tmp..733T,2018ApJ...859L...5M,kroupa2018does,2018MNRAS.480..473F,2018arXiv181207345E,2018arXiv181207346F,2019arXiv190102679M,2019MNRAS.484..510N,2019MNRAS.484..245L}. These two galaxies, NGC\,1052-DF2\footnote{
{Originally, this galaxy was 
first discovered by \citet{Fosbury1978} and also reported by \citet{2000A&AS..145..415K} and henceforth dubbed
[KKS2000]\,04, but here we follow the naming convention in recent literature.}
} \citep{2018Natur.555..629V},
  and NGC\,1052-DF4 \citep{2019arXiv190105973V} are 
very extended ($r_{eff}>1500$\,pc) and 
very dim {($mu_{eff,V}>25$\,mag arcsec$^{-2}$)} at a putative distance of NGC\,1052, though, {their distances are still} a matter of an active debate \citep{2018ApJ...864L..18V,2019MNRAS.tmp..733T}. Objects of this size and luminosity were first discovered in the Virgo Cluster by \citet{1984AJ.....89..919S} and 
{recent surveys of low surface brightness galaxies identified them in}
different galactic environments \citep[e.g. ][]{2015ApJ...809L..21M,2016ApJ...833..168M,2016A&A...590A..20V,2016MNRAS.463.1284O,2017MNRAS.470.1512W,2017A&A...608A.142V,2017MNRAS.468..703R,2017MNRAS.468.4039R,2015AstBu..70..379K,2018ApJ...855..142E,2018LeoI,2018ApJ...857..104G,2019MNRAS.485.1036M}. This class of galaxies has been rebranded as ultra diffuse galaxies (UDG, \citealt{2015ApJ...798L..45V}).

The apparent lack of dark matter of NGC\,1052-DF2 and NGC\,1052-DF4 makes these objects interesting and has raised discussions about their origin.  A primordial dwarf galaxy should always contain a dynamically significant portion of dark matter -- see, e.g. the dual dwarf theorem in \citet{2012PASA...29..395K} and its implications -- making the possible absence of dark matter in these galaxies a direct confrontation with the current cosmological paradigm.
 Several scenarios have been suggested, for example, that they were tidally stripped from their dark matter by a close-passage to NGC\,1052 \citep{2018MNRAS.480L.106O}, or that they are the direct results of a tidal interaction, i.e. Tidal Dwarf Galaxies (TDG, \citealt{2012ASSP...28..305D,2014MNRAS.440.1458D}).  
{{Mergers producing such TDGs} are expected to leave observable signatures in the form of several streams and shells around NGC\,1052 (and/or the other massive galaxies in the field) that can potentially be detected by deep imaging for several Gyr after the interaction.}
 
NGC\,1052  is an active radio galaxy consisting of two radio sources, a compact one, and an extended one in the form of a two-sided radio jet \citep{2003MNRAS.346..327T, 2004A&A...420..467K}. 
\citet{2005MNRAS.358..419P} {studied the recent history of NGC\,1052 and argued}  that the young central starburst had to start from  {gas enriched in} $\alpha$ elements. 
Such gas is usually found in massive spirals. This young central starburst has an age estimate of $\sim$1\,Gyr coinciding with the estimated age of the accretion of the  \hi\ clouds  found around NGC\,1052 \citep{1986AJ.....91..791V}. Its distance is estimated with surface brightness fluctuations (SBF) and measured to be $D=19$\,Mpc \citep{	2013AJ....146...86T}.
Another giant galaxy in the same field is NGC\,1042 -- a spiral galaxy separated by 15$\arcmin$ from NGC\,1052 to its south-east. There have been various distance estimations placing NGC\,1042 between 4 and 20\,Mpc \citep{1992ApJS...80..479T,2007A&A...465...71T,2008ApJ...676..184T,2016ApJ...823...85L}. A reconstruction of the velocity field of the galaxies in the local universe puts NGC\,1042 at a distance of 13\,Mpc \citep{2007A&A...465...71T}, which would mean that NGC\,1052 and NGC\,1042 are companions only in projection. However, they have similar velocities with $v=1376$\,km s$^{-1}$ for NGC\,1042 \citep{2009yCat.2294....0A} and $v=1510$\,km s$^{-1}$ for NGC\,1052 \citep{	2005MNRAS.356.1440D}. {Furthermore, in the extended field around NGC\,1052, there appears to be a plethora of giant galaxies sharing the same velocity (\citealt{1993A&AS..100...47G}, see also Section 5 and 6 of \citealt{2019MNRAS.tmp..733T}).}

In this work, we raise the question whether NGC\,1052-DF2 {and NGC\,1052-DF4 are} physically associated with the giant galaxies, as well as whether  tidal interactions can be observed between the galaxies. We have acquired deep images to recover the low-surface brightness features of the field around NGC\,1052 {and} probe the signatures of tidal features. 

\section{Observations and data processing}
{As part of the Halos and Environments of Nearby Galaxies (HERON) survey \citep{Rich2017IAU,Rich2019}, }
we have acquired deep images with the 0.7 meter Jeanne Rich telescope 
{\citep{Brosch2015,Rich2017IAU}} located at Lockwood Valley (CA, USA) between October and November 2018. The camera has {a pixel scale of 1.1 arcsec and a field} of view of {$\sim 1$} square degree, allowing us to cover NGC\,1052, NGC\,1042, NGC\,1052-DF2, and NGC\,1052-DF4 in a single image field.
In total, we {secured} {$245\times300$ second exposures, giving a total exposure of 20.4 hours. {The observations were conducted at an airmass between 1.4 and 1.8.}} {All observations were taken under photometric conditions when the moon was bellow the horizon.} {During the observations the sky brightness was between 21.7 to 22.0\,mag arcsec$^{-2}$ ($V$-band).}
 The reduction process was carried out by correcting bias and flat-fielding. For the creation of the flat, we mask the science images using Noisechisel \citep{2015ApJS..220....1A} and combine them   . Due to the presence of gradients, we use Noisechisel to obtain the sky of each image, which is subtracted from the already corrected images from bias and flat-fielding. Astrometry was obtained through the scamp package \citep{2006ASPC..351..112B}. Finally we combine all exposures with a 3$\sigma$ resistant mean.

The zero point was derived using APASS \citep{2014CoSka..43..518H}. To transform from our instrumental luminance $L$ magnitudes to the commonly used SDSS $r$ band, we applied a linear transformation using {25} standard stars from the APASS catalog. 
{We determine the surface brightness limits by measuring the standard deviation of blank pixels (pixels with no sources) in $10\times10$ arcsec boxes on the luminance $L$  image. The $3\sigma$ standard deviation -- transformed to the $r$-band using the linear photometric transformation -- is $\mu_{lim,r}= 28.5$\,mag arcsec$^{-2}$.
This limit} is comparable to the depth reached by other telescopes of modest aperture size \citep{2010AJ....140..962M,2015AstBu..70..379K,2016DGSAT,2017DGSAT}.

      \begin{figure*}[ht]
\includegraphics[width=18cm]{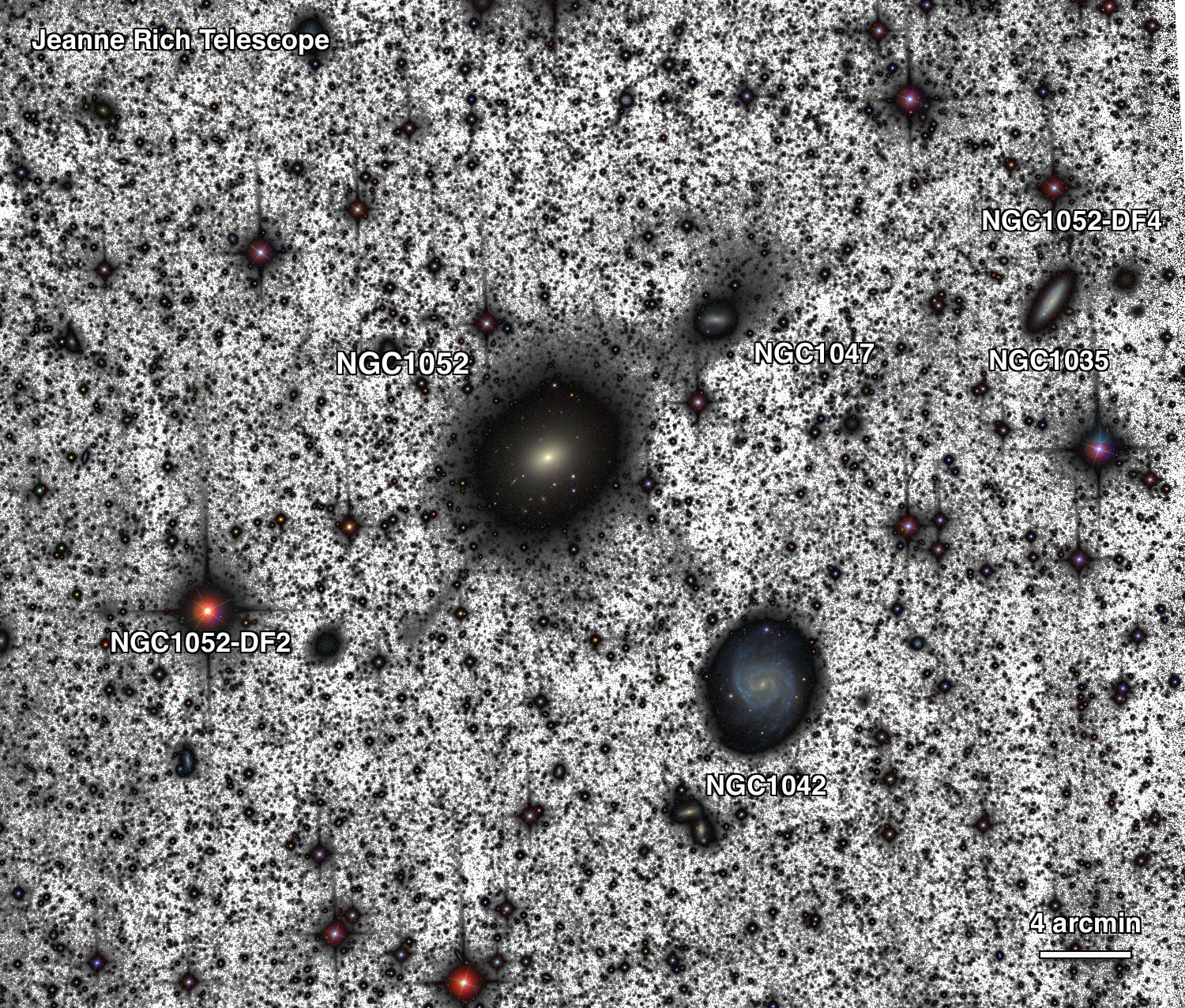}
\caption{{The NGC 1052 environment: stacked luminance (L) band  obtained with the Jeanne Rich telescope. True color images from the SDSS are overlayed. The field of view is 1\,deg$^2$. The indicated 4$\arcmin$ bar corresponds to 15\,kpc at 13\,Mpc or 22\,kpc at 19\,Mpc.
North is to the top, east to the left.}}
\label{field}
\end{figure*}

\section{Low-surface brightness features}
\begin{figure*}[ht]
\includegraphics[width=18cm]{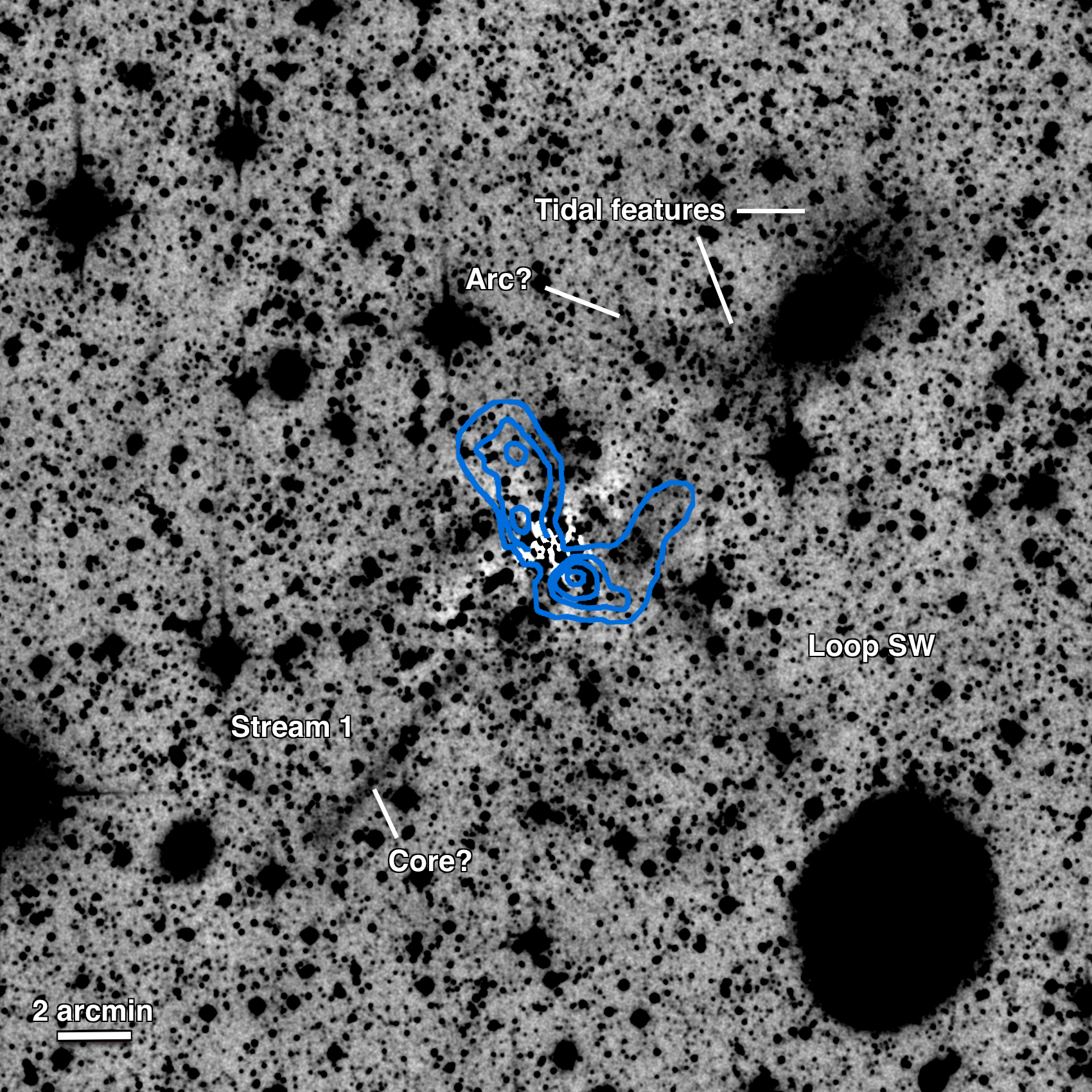}

\caption{The residual image after NGC\,1052 was subtracted {and convolved with a gaussian kernel}, revealing the low surface brightness features in its environment.
{The contours show the \hi\ distribution around NGC\,1052 adopted from \citet{1986AJ.....91..791V}.} 
}
\label{lsbs}
\end{figure*}

Several Low-surface brightness (LSB) features stand out in the field around NGC 1052 by merely having a glance at Figure \ref{field}.
We have modelled NGC\,1052 using IRAF's {\em ellipse}  algorithm \citep{1986SPIE..627..733T} after removing the faint objects and masking bright stars together with the companion galaxies. The model-subtracted image is presented 
in Figure \ref{lsbs}. 
Most prominently, there is a narrow stream coming from NGC\,1052 to the vicinity of
NGC\,1052-DF2 which we refer {to} as Stream 1. A loop is visible to the south-west (loop SW), close to NGC 1042 but apparently disconnected from it. 
The tidal tail as seen in \hi\ around NGC\,1052 extends towards NGC\,1047 (see Figure \ref{lsbs}). 
NGC\,1047 shows a clear tidal distortion/extension to the north-west, which is on the opposing side in respect to NGC\,1052. {NGC\,1047 also shows tidal features towards NGC\,1052, which could be a bridge}. {The outer isophotes of NGC\,1047 appear to be boxy.}
We also note an arc feature to the north{-west} of NGC\,1052.
While we find multiple tidal features around NGC\,1052, no convincing features are visible around NGC\,1042.

Stream 1 is a striking feature. It appears as a straight line coming from NGC\,1052 {in the direction of NGC\,1052-DF2}.
It stops 100 arcsec before the center of NGC\,1052-DF2.
The residual image of NGC 1052 shows that Stream 1 is extending towards the center of the bright elliptical galaxy. 

\begin{figure}[ht]
      \centering
\includegraphics[width=8.8cm]{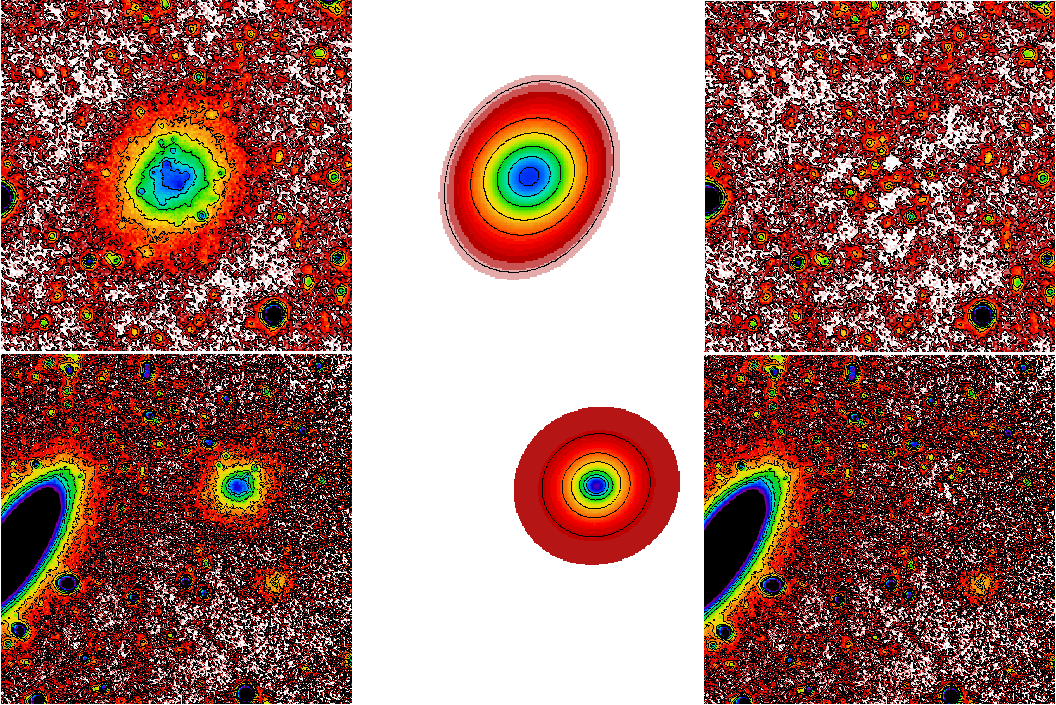}
\caption{The modelling of NGC\,1052-DF2 (top) and NGC\,1052-DF4 (bottom). Left: Isophotal levels on the original image; Middle: Model of the galaxies; Right: The residual image.} 
\label{fitting}
\end{figure}

To study the LSB features around NGC\,1052-DF2 and NGC\,1052-DF4 we have modelled and subtracted the two galaxies from the images (Figure \ref{fitting}). After the subtraction, no remaining features are left. While NGC\,1052-DF2 has a change of the Position Angle with respect to the radius, NGC\,1052-DF4 shows no sign of it.

\section{Discussion}
 NGC\,1052 resides in an environment where multiple interactions occurred in the past and are still on-going. 
  {The LSB features between NGC\,1052 and NGC\,1047 and the lopsided LSB feature on the counterside of NGC\,1047 undoubtedly show that NGC\,1052 and NGC\,1047 have an ongoing interaction and are therefore at the same distance. The boxy appearance of the outer isophotes of NGC\,1047 further indicates tidal disturbance.}
 NGC\,1052 also contains an irregularly shaped distribution of \hi\ \citep{1986AJ.....91..791V}, with what appears to be a tidal arm extending towards the south-west. 
\citet{1986AJ.....91..791V} argued
that the \hi\ has been acquired 1\,Gyr ago. In addition, \citet{2003MNRAS.346..327T} argued that NGC\,1052 is a restarted radio source. The extended radio source is likely much older than the compact radio source at the nucleus, and these sources could have been created during different events.

The relative position of Stream 1 and NGC\,1052-DF2 {may suggest} 
that the galaxy is a TDG formed at a tip of a tidal arm \citep{2006A&A...456..481B}.  Indeed old TDGs have the structural properties (central surface brightness, large effective radius, lack of dark matter content) of NGC\,1052-DF2 and NGC\,1052-DF4 \citep{2014MNRAS.440.1458D}.
While the hypothesis of a young TDG is falsified by the stellar age of NGC\,1052-DF2 (9\,Gyr, \citealt{2018arXiv181207346F}), one could argue that it is an old TDG, formed during an interaction 9 Gyr ago. This could solve the dark matter-deficiency issue and would be consistent with the location of NGC\,1052-\,DF2 on the metallicity-stellar mass relation (see discussion in \citealt{2018arXiv181207346F}). However, tidal features, such as Stream 1 should disappear in a few Gyrs, especially if the galaxy suffered posterior mergers. 
{Namely,} some of the most extended tidal arms are around 100\,kpc long and the free fall time from such a distance to NGC\,1052 is about 1-2\,Gyr, i.e. substantially less than the estimated age of NGC\,1052-DF2. 
The linear morphology of Stream 1 is not expected if the stream had made several orbits around the host galaxy. Hence we can conclude that the alignment between this stream and NGC 1052-DF2 is pure chance and not linked with any TDG scenario.
In addition, Stream 1 will not be the stripped tail of NGC\,1052-DF2, because in such a case we would  expect the tidal tail and the progenitor being connected (see, e.g. Figure 10  of \citealt{2018A&A...615A..96M}). The morphology of the observed tidal features does not agree with the morphology of the tidal features in the simulation by \citet{2018MNRAS.480L.106O}. 
Alternatively and perhaps more likely, Stream 1 could originate from another stripped dwarf galaxy satellite.
We note an unresolved source on the axis of the stream which is a candidate for the stripped core of its progenitor (02:41:25.3, $-$08:22:11.6). 
All of these suggest that this is likely a chance alignment.  This is consistent with  the subtracted model of NGC\,1052-DF2 (Figure \ref{fitting})  revealing no clear signs of tidal features.

Turning to NGC\,1042, also the absence of LSB features contains valuable information.  While its systemic velocity ($v=1376$\,km s$^{-1}$; \citealt{2009yCat.2294....0A}) is consistent with the velocity of the NGC\,1052 group ($v=1425\pm111$\,km s$^{-1}$; \citealt{2018Natur.555..629V}), i.e. a distance of $\approx20$\,Mpc, the redshift-independent distance estimators favor a closer value (13\,Mpc \citealt{2007A&A...465...71T}). As with NGC\,1047, a visible tidal interaction would, without a doubt, put NGC\,1042 at the NGC\,1052 group.  There is a loop which is in the direction of NGC\,1042 but its morphology is rather consistent with a stream associated to another disrupted satellite orbiting around NGC\,1052. No clear tidal features are directly connected to NGC\,1042.  This can either mean that NGC\,1042 is within the group or has just arrived in the group, and no visible interaction has yet taken place\footnote{For example, the  systemic velocities for Cen\,A and M\,83 are consistent with each other, but their separation puts them more than 1.1\,Mpc away \citep{2013AJ....145..101K}, and no connecting LSB features have been found.}, or that NGC\,1042 is indeed only associated to NGC\,1052 in projection and is, in reality, closer to us. 
\citet{2019MNRAS.tmp..733T} argue that the properties of NGC\,1052-DF2 and NGC\,1052-DF4 would match with other known dwarf galaxies if they were closer to us, for instance  at the distance of 13\,Mpc.  {On the other hand, NGC\,1042's spiral structure itself shows sharp changes in pitch angle and loosely wound outer spiral arms which are quite flocculent, which is reminiscent of  spirals in the early time steps of merger simulations, not having  quite managed to form elongated tidal tails yet.} 

{The remaining giant galaxy to be discussed in the observed field is the spiral galaxy NGC\,1035. While it also has a systemic velocity consistent with the NGC\,1052 group ($v=1249$\,km s$^{-1}$; \citealt{2004MNRAS.350.1195M}), its latest distance estimate -- based on Tully-Fisher measurements -- is $14\pm2.9$\,Mpc \citep{2014MNRAS.444..527S}. This places it in the foreground and is consistent with an absence of tidal features associated with this galaxy.}

\section{Conclusions}

We aim to clarify the origin of the alleged low dark matter content of NGC\,1052-DF2 and NGC\,1052-DF4 by deep imaging.   Deep imaging can, in principle, uncover tidal features whose morphology would reveal the origin of NGC\,1052-DF2 and NGC\,1052-DF4 either as tidal dwarf galaxies or a tidally dark-matter stripped primordial dwarf. We indeed found signs of many galaxy interactions in the group. Notably, there is an ongoing tidal interaction between NGC\,1052 and NGC\,1047.
Furthermore, around NGC\,1052 we find a loop in the direction of NGC\,1042 and a stream in the direction of NGC\,1052-DF2, but none of them are directly connected to these two galaxies.
The features around NGC\,1052 are  probably much younger (1-3\,Gyr) and are inconsistent with the estimated age of NGC\,1052-DF2 (9\,Gyr). 
We see no tidal features 
around  NGC\,1052-DF2 and NGC\,1052-DF4, making a recent tidal stripping scenario unlikely.
The tidal dwarf galaxy scenario at high redshift cannot be probed by our data as the tidal features might have been erased and are not detectable today.

We further found no convincing tidal features around the spiral galaxy NGC\,1042. This indicates that either it is at a larger separation within the NGC\,1052 group{, e.g. just falling in}, or that its association is only in projection and in reality, NGC\,1042 is closer to us. {Their final association (physical or projected) will be only firmly established once we have accurate distance estimations to both objects.}

\begin{acknowledgements}
{The authors thank the referee for the timely and helpful report.} The authors {also} thank Eric Emsellem, Anita Zanella, Remco van der Burg, Marina Rejkuba, Michael Hilker, Federico Lelli, Marcel Pawlowski, and Noam Libeskind for interesting discussions and comments.
 OM is grateful to the Swiss National Science Foundation for financial support. R.M.R. acknowledges the Polaris Observatory Association for hosting the Jeanne Rich telescope, Ian Kearns-Brown for supporting operations, and Parkes Whipple for assistance with data reduction. M.K.Y. acknowledges the financial support from the Scientific and Technological Research Council of Turkey (TUBITAK) under the postdoc fellowship programme 1059B191701226.
 I.T. acknowledges financial support from the European Union's Horizon 2020 research and innovation programme under Marie Sklodowska-Curie grant agreement No 721463 to the SUNDIAL ITN network. This research has been partly supported by the Spanish Ministry of Economy and Competitiveness (MINECO) under grants AYA2016-77237-C3-1-P.
\end{acknowledgements}

\bibliographystyle{aa}
\bibliography{bibliographie}

\end{document}